\documentclass[10pt,pra,twocolumn,superscriptaddress,showpacs,floatfix]{revtex4}

\usepackage{graphicx}
\usepackage{amssymb}
\usepackage{times}
\usepackage{amsmath}
\usepackage{amsthm}
\usepackage{epstopdf}

\usepackage{hyperref}

\usepackage{color}
\begin{document}
	
\title{Quantum hacking: Induced-photorefraction attack on a practical continuous-variable quantum key distribution system}

\author{Yiliang Wang}
\affiliation
{School of Computer Science, Northwestern Polytechnical University, Xi'an 710129, Shaanxi, China
}
\author{Yi Zheng}\thanks{yizheng@nwpu.edu.cn}
\affiliation
{School of Computer Science, Northwestern Polytechnical University, Xi'an 710129, Shaanxi, China
}
\author{Chenlei Fang}
\affiliation
{School of Computer Science, Northwestern Polytechnical University, Xi'an 710129, Shaanxi, China
}
\author{Haobin Shi}
\affiliation
{School of Computer Science, Northwestern Polytechnical University, Xi'an 710129, Shaanxi, China
}
\author{Wei Pan}
\affiliation
{School of Computer Science, Northwestern Polytechnical University, Xi'an 710129, Shaanxi, China
}

\begin{abstract}
    We explore a new security loophole in a practical continuous-variable quantum key distribution (CVQKD) system, which is opened by the photorefractive effect of lithium niobate-based (LN-based) modulators. By exploiting this loophole, we propose a quantum hacking strategy, i.e., the induced-photorefraction attack, which utilizes the induced photorefraction on the LN-based modulators to hide the classical intercept-resend attack. Specifically, we show that the induced photorefraction can bias the response curve of the LN-based modulator, which will affect the intensity of the modulated signal. Based on the investigation of the channel parameter estimation under above influence, we further analyze the secret key rate of the practical CVQKD system. The simulation results indicate that the communication parties will overestimate the secret key rate, which reveals that Eve can actively open the above loophole by launching the induced-photorefraction attack to successfully obtain the secret key information. To defend against this attack, we can use a random monitoring scheme for modulation variance to determine this attack, and use an improving optical power limiter to effectively mitigate the irradiation beam. Apart from these countermeasures, we also propose using the Sagnac-based IM to stabilize the practical CVQKD system, which can minimize the above effects.
\end{abstract}


\pacs{03.67.Hk, 03.67.-a, 03.67.Dd}
\maketitle

\section{Introduction}\label{sec1}
Quantum key distribution (QKD) technology allows two authenticated communication parties, Alice and Bob, to securely exchange secret keys over a quantum channel \cite{r1}, even if there is a potential eavesdropper, Eve. The fundamental laws of quantum physics ensure its theoretical unconditional security \cite{r2,r3}. In recent years, as one of the most mature technologies in the field of quantum cryptography, QKD has made tremendous development in both theory and experiment \cite{r4,r5,r6,r7}. There are two promising approaches to perform quantum key distribution, which are discrete-variable QKD (DVQKD) and continuous-variable QKD (CVQKD). Different from the DVQKD protocol, the CVQKD protocol encodes secret key information on the light field quadratures \cite{r8}, which can be measured by homodyne detectors instead of the single-photon counters. 

Gaussian-modulated coherent state (GMCS) scheme \cite{r9} is one of the most favorable CVQKD schemes, which has been fully proven to be secure against both collective attacks and coherent attacks \cite{r10,r11,r12}. However, in a practical GMCS CVQKD system, the deviations between the realistic devices and the theoretical model may open security loopholes for Eve to successfully obtain secret key information \cite{r13,r14}. So far, a number of quantum hacking attacks have been proposed, such as wavelength attack \cite{r15,r16}, saturation attack \cite{r17}, detector blinding attack \cite{r18}, finite sampling bandwidth effects \cite{r19}, polarization attack \cite{r20}, local oscillator (LO) fluctuation attack \cite{r21}, LO calibration attack \cite{r22}, laser damage attack against optical attenuator \cite{r23}, and extern magnetic fields attack against optical amplifier \cite{r24}. Correspondingly, several countermeasures have been proposed by researchers. For example, a local LO (LLO) scheme \cite{r25,r26,r27,r28} has been designed to close the security loopholes related to LO. Furthermore, the proposed continuous-variable measurement-device-independent quantum key distribution (CV-MDI-QKD) system \cite{r29,r30,r31} can remove all security loopholes in detection side. Therefore, compared to the detection side, the source side is the more vulnerable part and its practical security should receive more attention.

In a practical GMCS CVQKD system, lithium niobate-based (LN-based) Mach-Zehnder (MZ) modulators are widely used to prepare appropriate coherent states. Since the lithium niobate is a kind of photorefractive material, external irradiation beam will affect its refractive index distribution. Based on this property, we propose an induced-photorefraction attack in a practical GMCS CVQKD system. Specifically, we first introduce the working principle of LN-based MZ modulators and the mechanism of photorefractive effect (PE). Subsequently, we describe the process of the induced photorefraction on a LN-based MZ modulator and analyze its effects on the practical GMCS CVQKD system. We find that the response curve of a LN-based MZ modulator will drift under above effects, which can lead to amplification of the intensity of the optical signal sent by Alice. Then, we investigate the quantum channel parameter estimation under the effect of the induced photorefraction. The result shows that the channel excess noise will be underestimated by Alice and Bob, which creates conditions for Eve to perform the full intercept-resend attack. By combining the induced photorefraction and the intercept-resend attack, we design the schemes of the induced-photorefraction attack (IPA). In addition, we simulate the relationship between the secret key rate and the transmission distance under the induced-photorefraction attack.  The simulation results show that the secret key rate is overestimated by Alice and Bob, which will open a security loophole for Eve to obtain secret key information without being detected. To resist the induced-photorefraction attack, corresponding countermeasures are proposed. Firstly, we can use a random monitoring scheme of modulation variance to determine the intensity of the induced-photorefraction attack. Secondly, an improving optical power limiter is proposed to effectively limit the irradiation beam and detect Eve’s attacks. Lastly, we suggest using the Sagnac-based IM with natural stability instead of the LN-based MZ IM to enhance the security of the practical CVQKD system.

This paper is organized as follows. In Sec. \ref{sec2}, we focus on the working principle of LN-based MZ modulators and the mechanism of PE, and analyze the effect of the induced photorefraction on LN-based MZ modulators. Then, in Sec. \ref{sec3}, we introduce the schemes of an induced-photorefraction attack and investigate its effects on the channel parameter estimation and secret key rate of the CVQKD system. And in Sec. \ref{sec4}, different countermeasures are proposed to close the security loophole induced by the induced-photorefraction attack. Finally, we draw the conclusion in Sec. \ref{sec5}.

\section{CVQKD system under the effect of the induced photorefraction}\label{sec2}
\subsection{The working principle of LN-based MZ modulator}\label{subsec2-1}
\begin{figure}[!h]\center
    \centering
    \resizebox{8cm}{!}{
        \includegraphics{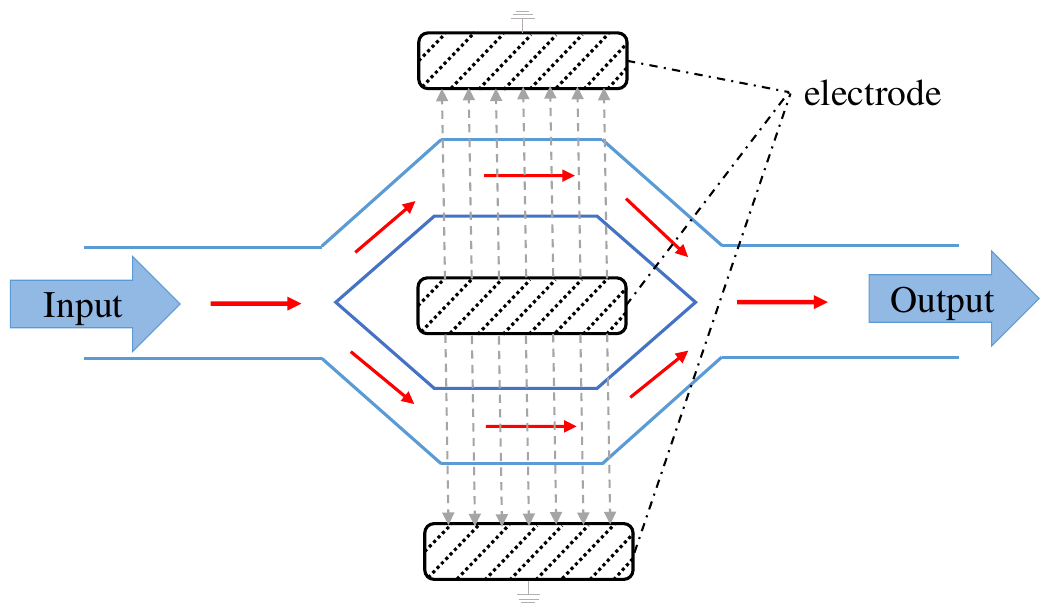}}
    \caption{Structure of a LN-based MZ modulator.}
    \label{FIG3}
\end{figure}

Lithium niobate-based Mach-Zehnder (LN-based MZ) modulators, such as IM and MZ VOA, are widely used to prepare quantum-level coherent states carrying the secret key information in practical CVQKD systems (See Appendix \ref{appendixA} for more details about the CVQKD system). Specifically, the LN-based MZ modulator first modulates the phase by changing the refractive index of the LN waveguide, and then indirectly realizes the intensity modulation of signal light through the interference of light. The structure of the LN-based MZ modulator is shown in Fig. \ref{FIG3}.

It is obvious that the LN-based MZ modulator mainly consists of input waveguide, beam splitter structure, phase modulation and output waveguide. And the phase modulation is the core component, which consists of three electrodes and LN waveguides. The high electrode is located in the center and the grounded electrodes are on both sides. Therefore, the two arms of the modulator are in electric fields of equal intensity but opposite directions. When the signal light enters the modulator, it will first be divided into two identical beams by the beam splitter and enter the two arms of the modulator respectively. Then, due to the effect of the electric field, the signal light in the two arms will have a phase drift of same magnitude but opposite directions respectively. Finally, these two beams of signal light interfere and output. The above steps realize the intensity modulation of the signal light.

Here, the wave function of the input signal light of the modulator can be expressed as
\begin{equation}\label{e1}
    A_{in}(t)=A_{0}exp(i\omega_{0}t),\\
\end{equation}
where $A_0$ and $\omega_{0}t$ are the amplitude and phase of the input signal light, respectively. After passing through the beam splitter, the signal light entering the two arms of the modulator can be represented as
\begin{equation}\label{e2}
    A_{1}(t)=A_{2}(t)=\frac{1}{\sqrt{2}}A_{0}exp(i\omega_{0}t).\\
\end{equation}
Subsequently, the signal light after the phase modulation in the two arms of modulator can be expressed as
\begin{equation}\label{e3}
    \begin{split}
        A_{1}^{\prime}(t)&=\frac{1}{\sqrt{2}}A_{0}exp[i(\omega_{0}t+\varphi_{1})],\cr
        A_{2}^{\prime}(t)&=\frac{1}{\sqrt{2}}A_{0}exp[i(\omega_{0}t+\varphi_{2})],
    \end{split}
\end{equation}
where $\varphi_1$ and $\varphi_2$ are the phase drifts under the effect of the electric field. Finally, after the interference of the signal light in the two arms of modulator, the wave function of the output signal light can be represented as
\begin{equation}\label{e4}
    \begin{split}
        A_{out}(t)&=\frac{1}{\sqrt{2}}[A_{1}^{\prime}(t)+A_{2}^{\prime}(t)]\cr
        &=\frac{1}{2}A_{0}\{exp[i(\omega_{0}t+\varphi_{1})]+exp[i(\omega_{0}t+\varphi_{2})]\}\cr
        &=A_{0}\cos\frac{\varphi_{1}-\varphi_{2}}{2}exp\Big[i({\omega}_{0}t+\frac{\varphi_{1}+\varphi_{2}}{2})\Big].
    \end{split}
\end{equation}
Since the phase drift $\varphi_1$ and $\varphi_2$ are of the same magnitude but opposite directions, i.e., $\varphi_1+\varphi_2=0$, we can get that
\begin{equation}\label{e5}
    A_{out}(t)=A_{0}\cos\frac{\varphi_{1}-\varphi_{2}}{2}exp(i{\omega}_{0}t).\\
\end{equation}

\begin{figure}[!h]\center
    \centering
    \resizebox{8cm}{!}{
        \includegraphics{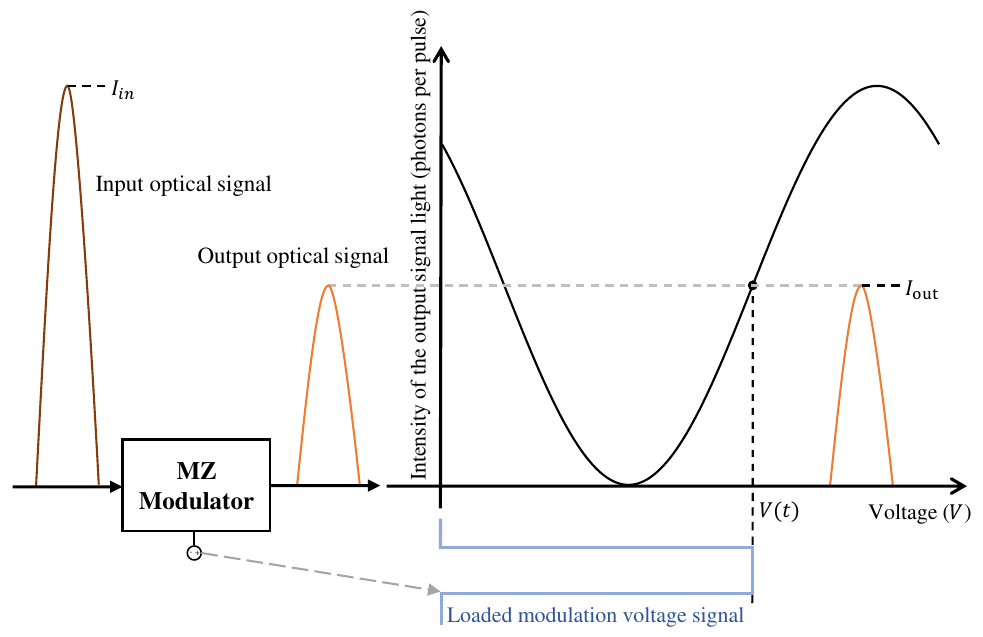}}
    \caption{Working principle diagram of a LN-based MZ modulator.}
    \label{FIG4}
\end{figure}

Moreover, the intensity $I$ and amplitude $A$ of the signal light obey the following relation:
\begin{equation}\label{e6}
    I\propto{|A|}^2.\\
\end{equation}
Therefore, the intensity of the output signal light can be calculated as 
\begin{equation}\label{e7}
    \begin{split}
        I_{out}&=I_{in}{\cos}^2\Big(\frac{\varphi_{1}-\varphi_{2}}{2}\Big)\cr
        &=\frac{1}{2}I_{in}[1+\cos(\varphi_{1}-\varphi_{2})]\cr
        &=\frac{1}{2}I_{in}[1+\cos(2\varphi_{1})],
    \end{split}
\end{equation}
where $I_{in}$ is the intensity of the input signal light. Since the modulation phase is linearly related to the loaded voltage, we can obtain that
\begin{equation}\label{e7-1}
    \varphi_{1}=\frac{\pi}{V_{\pi}}V,\\
\end{equation}
where $V_{\pi}$ is the half-wave voltage, $V=V_{bias}+V(t)$, $V_{bias}$ is the bias voltage, and $V(t)$ is the loaded modulation voltage. Then, Eq (\ref{e7}) should be written as
\begin{equation}\label{e7-2}
    \begin{split}
        I_{out}&=\frac{1}{2}I_{in}\big[1+\cos\big(2\frac{\pi}{V_{\pi}}V\big)\big]\cr
        &=\frac{1}{2}I_{in}\big[1+\cos\big(2\frac{\pi}{V_{\pi}}[V_{bias}+V(t)]\big)\big]\cr
        &=\frac{1}{2}I_{in}\big[1+\cos[2\frac{\pi}{V_{\pi}}V_{bias}+2\frac{\pi}{V_{\pi}}V(t)]\big]\cr
        &=\frac{1}{2}I_{in}\{1+\cos[\varphi_{0}+kV(t)]\},
    \end{split}
\end{equation}
where $\varphi_0=\frac{2\pi}{V_{\pi}}V_{bias}$ is the initial inherent phase, $k=\frac{2\pi}{V_{\pi}}$ is a constant coefficient. Based on above analysis, we depict the working principle diagram of the LN-based MZ modulator (see Fig. \ref{FIG4}). It is observed that the intensity of the output signal light of LN-based MZ modulator is $I_{out}$ when the loaded modulation voltage is $V(t)$. Therefore, in the modulation process of a practical CVQKD system, Alice and Bob can change the intensity of the output signal light by controlling the value of the loaded modulation voltage.

\subsection{The effect of the induced photorefraction}\label{subsec2-2}
The photorefractive effect (PE) process \cite{r32} can be described as follows: photogenerated charge pairs are generated under the illumination of the irradiation beam; the charges redistribute driven by the carrier concentration distribution, electric field or photovoltaic effect; the inhomogeneous charge distributions create a space-charge field, which affects the refractive index of the lithium niobate waveguide through the Pockels effect. Based on the above characters of PE, the refractive index of the lithium niobate waveguide can be changed by the irradiation beam, which may be utilized by Eve to perform induced photorefraction to affect the modulation results of the LN-based MZ modulator. Specifically, the irradiation beam is first inversely injected into the LN-based MZ modulator. Subsequently, due to the wavelength dependence of the beam splitter structure, the intensity of the irradiation beam into the two arms of the modulator is different. Furthermore, the fabrication error and the applied electric fields are also different for the two arms of the modulator. Therefore, the PE on these two arms is inconsistent, which results a PE phase difference ${\Delta\theta}_{P\!E}$ in the LN-based MZ modulator. Notably, the PE process practically affects both the phase of the signal light and the beam-splitting ratio (BSR) of the beam splitter structure. However, the effect of PE process on BSR is almost negligible when the power of the irradiation beam is low ($<$100mW) \cite{r33}. Therefore, in this paper, we focus on the variation of the phase of the signal light. Next, we will further analyze the effect of induced photorefraction on the intensity of the output signal light of the LN-based MZ modulator.
\begin{figure}[!h]\center
\centering
\resizebox{8cm}{!}{
    \includegraphics{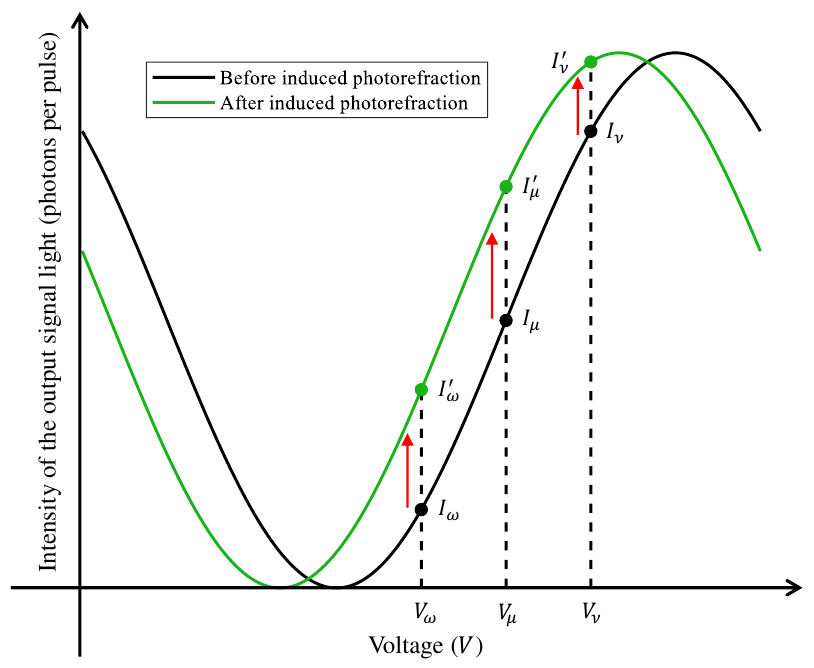}}
\caption{Response curve of a LN-based MZ modulator under the effect of induced photorefraction.}
\label{FIG5}
\end{figure}

According to Eq. (\ref{e7-2}) and above analysis, the intensity of the output signal light of the LN-based MZ modulator under the effect of induced photorefraction should be expressed as
\begin{equation}\label{e8}
    I_{out}^{\prime}=\frac{1}{2}I_{in}\{1+\cos[\varphi_{0}+kV(t)+\Delta\theta_{P\!E}]\}.\\
\end{equation}
Figure \ref{FIG5} shows the effect of induced photorefraction on the response curve of a LN-based MZ modulator. It is worth noting that we assume the value of ${\Delta\theta}_{P\!E}$ can be completely controlled by Eve. The black and green lines denote the response curve of a LN-based MZ modulator before and after executing the induced photorefraction, respectively. To further analyze the impact of the induced photorefraction on a practical CVQKD system, we assume that the modulation voltage of the LN-based MZ IM\_1 used in the pulse modulation in Fig. \ref{FIG1} is $V_{\nu}$, the modulation voltage of the LN-based MZ IM\_2 used in the Gaussian modulation is $V_{\mu}$, and the modulation voltage of the LN-based MZ VOA used in the signal optical path is $V_{\omega}$. As shown in Fig. \ref{FIG5}, before executing the induced photorefraction, the intensities of the output signal light of the IM\_1, IM\_2 and MZ VOA are $I_{\nu}$, $I_{\mu}$ and $I_{\omega}$, respectively. However, under the effect of the induced photorefraction, the practical intensities of the output signal light of the IM\_1, IM\_2 and MZ VOA become $I_{\nu}^{\prime}$, $I_{\mu}^{\prime}$ and $I_{\omega}^{\prime}$, respectively. Here, we assume that
\begin{equation}\label{e9}
    \begin{split}
        I_{\nu}^{\prime}&=M_{1}I_{\nu} \quad(M_1>1),\\
        I_{\mu}^{\prime}&=M_{2}I_{\mu} \quad(M_2>1),\\
        I_{\omega}^{\prime}&=M_{3}I_{\omega} \quad(M_3>1).
    \end{split}
\end{equation}

\begin{figure}[!h]\center
\centering
\resizebox{8cm}{!}{
    \includegraphics{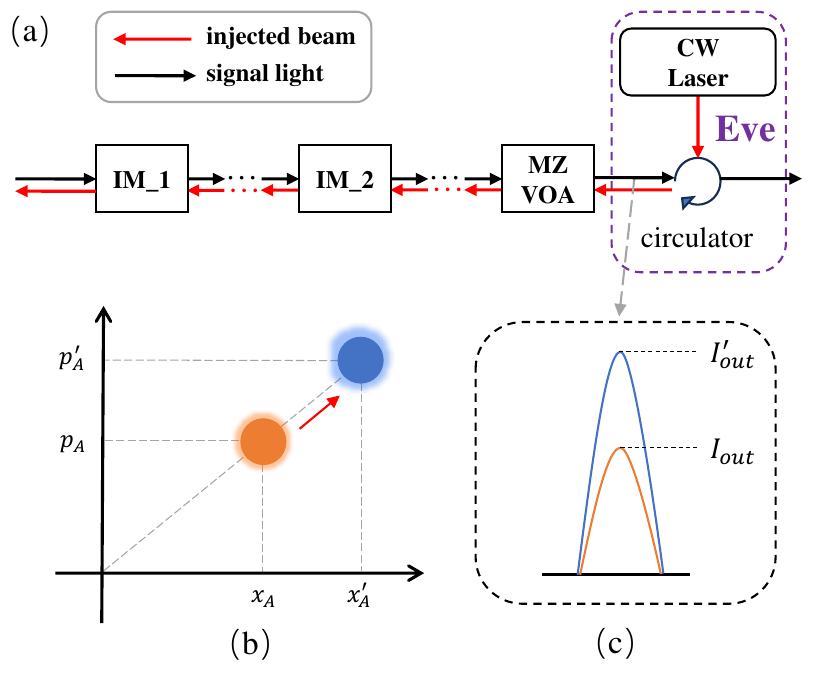}}
\caption{Effects of the induced photorefraction on a practical CVQKD system. (a) Simplified diagram of implementing induced photorefraction on a practical CVQKD system. (b) Expression of the transmitted Gaussian-modulated coherent state in the phase space under above influence. (c) Schematic diagram of output signal light intensity before and after executing the induced photorefraction. $I_{out}^{\prime}$, the intensity of the output signal light of MZ VOA under the effect of the induced photorefraction.}
\label{FIG6}
\end{figure}

Figure \ref{FIG6} shows the effects of the induced photorefraction on a practical CVQKD system. Based on the above analysis of LN-based MZ devices under the influence of the induced photorefraction, we find that the intensities of the output signal light for IM\_1, IM\_2 and MZ VOA will be amplified by $M_1$, $M_2$ and $M_3$ times, respectively. Therefore, when considering the above devices simultaneously in a practical CVQKD system, the intensity of the output signal light of MZ VOA will also be amplified under the effect of the induced photorefraction (see Fig. \ref{FIG6} (c)), which can be represented as
\begin{equation}\label{e10}
    I_{out}^{\prime}=MI_{out} \quad(M>1),\\
\end{equation}
where $M=\prod_{i=1}^{3} M_i$ is defined as the impact factor, which reflects the intensity of the induced photorefraction, and $I_{out}$ is the intensity of the output signal light of MZ VOA without the induced photorefraction. It is worth noting that under the effect of the induced photorefraction, $M_i>1$ $(i=1,2,3)$ is the worst case. In fact, the intensities of the output signal light of the LN-based MZ modulators may not be amplified at the same time, but as long as $M>1$, the induced photorefraction may be utilized by Eve to threaten the practical security of the CVQKD system.

In the phase space, the output signal light of MZ VOA can be expressed as
\begin{equation}\label{e11}
    \begin{split}
        |\alpha_{A}\rangle=&|\alpha_{A}|e^{i\theta}=x_{A}+ip_{A},\\
        x_{A}=&|\alpha_{A}|\cos\theta,\\
        p_{A}=&|\alpha_{A}|\sin\theta,
    \end{split}
\end{equation}
where $|\alpha_{A}|$ and $\theta$ are the amplitude and phase of the transmitted Gaussian-modulated coherent states, and the $x_A$ and $p_A$ are the quadrature variables of coherent states. Correspondingly, the $x_A$ and $p_A$ will become $x_{A}^{\prime}$ and $p_{A}^{\prime}$ under the effect of the induced photorefraction (see Fig. \ref{FIG6} (b)). Moreover, the variance $V_A$ of the quadrature $x_A$ or $p_A$ is proportional to the mean photon number of the signal light.

According to Eqs. (\ref{e6}), (\ref{e10}), (\ref{e11}), and above analysis, we can obtain that
\begin{equation}\label{e12}
    \begin{split}
        x_{A}^{\prime}&=\sqrt{M}x_{A},\\
        p_{A}^{\prime}&=\sqrt{M}p_{A},\\
        V_{A}^{\prime}&=MV_{A},
    \end{split}
\end{equation}
where $x_{A}^{\prime}$ and $p_{A}^{\prime}$ are the quadrature variables of coherent states under the effect of the induced photorefraction, and $V_{A}^{\prime}$ is the variance of the quadrature $x_{A}^{\prime}$ or $p_{A}^{\prime}$. 

It is noteworthy that the intensity of LO light will also be attenuated to a suitable value by the MZ VOA in the LO optical path. Therefore, the induced photorefraction also has an effect on the attenuation of LO light. Fortunately, due to the calibrated linear relation between the variance of shot noise and the intensity of LO light, the real-time shot-noise measurement technique can be used to eliminate the above effect by monitoring the intensity of LO light \cite{r23}. 

\section{Security analysis}\label{sec3}
\subsection{Parameter estimation under the induced photorefraction}\label{subsec3-1}
After the balanced homodyne detection, Alice and Bob will share a group of correlated Gaussian data $X=\left\lbrace(x_{A_i},x_{B_i})|i=1,2,\ldots,N\right\rbrace$ or $P=\left\lbrace(p_{A_i},p_{B_i})|i=1,2,\ldots,N\right\rbrace$, where $N$ is the total number of the received pulses. In a CVQKD system, the quantum channel is assumed to be a linear model \cite{r17,r20,r24}, which can be represented as
\begin{equation}\label{e13}
    x_{B}=tx_{A}+z,\\
\end{equation}
where $t=\sqrt{\eta T}$ and vector $z$ satisfies a centered Gaussian distribution with variance $\sigma^2=\eta T\xi+N_0+V_{el}$. Here, $\xi=\varepsilon N_0$, $V_{el}=v_{el}N_0$, and $N_0$ is the variance of the shot noise.

According to Eq. (\ref{e13}), we can obtain the following relationships between $x_A$ and $x_B$:
\begin{equation}\label{e14}
    \begin{split}
        V_{A}=Var&(x_A)=\langle {x_A^2}\rangle,\\
        V_{B}=Var(x_B)=\langle {x_B^2}\rangle&=\eta TV_A+\eta T\xi+N_0+V_{el},\\
        Cov(x_A,x_B)&=\langle x_Ax_B\rangle=\sqrt{\eta T}V_A.
    \end{split}
\end{equation}
Here, similar equations could be derived for the other quadrature variables $p_A$ and $p_B$. It should be noted that the parameters $\eta$ and $v_{el}$ will be calibrated before communication. Furthermore, Alice and Bob will randomly extract $m$ $(m<N)$ pairs data from the $X$ or $P$ to estimate the channel parameters $T$ and $\xi$, which can be deduced from Eq. (\ref{e14}):
\begin{equation}\label{e15}
    \begin{split}
        T&=\frac{{Cov(x_A,x_B)}^2}{\eta V_A^2},\\
        \xi&=\frac{V_B-N_0-V_{el}}{\eta T}-V_A.
    \end{split}
\end{equation}

Considering the effect of the induced photorefraction, the Eqs. (\ref{e13}), (\ref{e14}),  and (\ref{e15}) should be written as
\begin{equation}\label{e16}
    \begin{split}
        x_{B}^{\prime}&=t^{\prime}x_{A}^{\prime}+z^{\prime},\\
        V_{A}^{\prime}&=V\!ar(x_A^{\prime})=\langle {x_A^{\prime}}^2\rangle,\\
        V_{B}^{\prime}=V\!ar(x_B^{\prime})=\langle {x_B^{\prime}}^2\rangle&=\eta T_{pra}V_A^{\prime}+\eta T_{pra}\xi_{pra}+N_0+V_{el},\\
        Cov(x_A^{\prime},x_B^{\prime})&=\langle x_A^{\prime}x_B^{\prime}\rangle=\sqrt{\eta T_{pra}}V_A^{\prime},\\
        T_{pra}&=\frac{{Cov(x_A^{\prime},x_B^{\prime})}^2}{\eta {V_A^{\prime}}^2},\\
        \xi_{pra}&=\frac{V_B^{\prime}-N_0-V_{el}}{\eta T_{pra}}-V_A^{\prime},
    \end{split}
\end{equation}
where $x_{A}^{\prime}$ and $V_{A}^{\prime}$ are mentioned in Eq. (\ref{e12}), $t^{\prime}=\sqrt{\eta T_{pra}}$, ${z^{\prime}}\sim N(0,\eta T_{pra}{\xi}_{pra}+N_0+V_{el})$, $T_{pra}$ is regarded as the practical value of parameter $T$, and $\xi_{pra}$ is regarded as the practical value of parameter $\xi$. 

However, when Alice and Bob are not aware of the induced photorefraction, they still use $x_A$ to estimate the parameters $T$ and $\xi$ for a CVQKD system. Therefore, the estimated values of parameters $T$ and $\xi$ can be expressed as
\begin{equation}\label{e17}
    \begin{split}
        T_{est}&=\frac{{Cov(x_A,x_B^{\prime})}^2}{\eta V_A^2},\\
        \xi_{est}&=\frac{V_B^{\prime}-N_0-V_{el}}{\eta T_{est}}-V_A.
    \end{split}
\end{equation}

According to Eqs. (\ref{e12}), (\ref{e16}), and (\ref{e17}), we can obtain
\begin{equation}\label{e18}
    \begin{split}
        T_{est}&=MT_{pra},\\
        \xi_{est}&=\frac{\xi_{pra}}{M}.
    \end{split}
\end{equation}
Expressed in shot-noise units, the estimated excess noise can be written as
\begin{equation}\label{e19}
    {\varepsilon}_{est}=\frac{{\varepsilon}_{pra}}{M} \qquad(M>1).
\end{equation}
\begin{figure}[!h]\center
\centering
\resizebox{8cm}{!}{
    \includegraphics{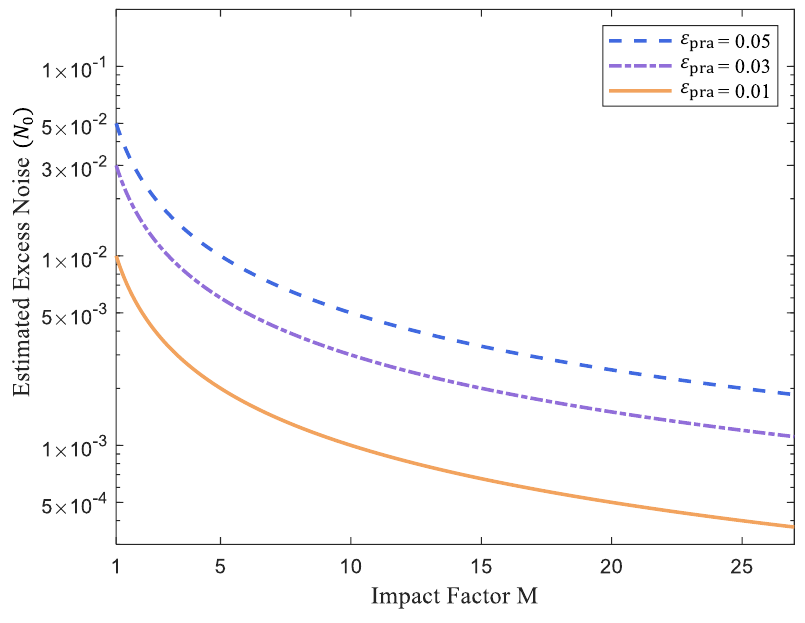}}
\caption{Estimated excess noise as a function of the impact factor $M$ under different values of the practical excess noise.}
\label{FIG7}
\end{figure}

Figure \ref{FIG7} depicts the relationship between the estimated excess noise $\varepsilon_{est}$ and the impact factor $M$ under different practical excess noises (i.e., $\varepsilon_{pra}\!=\!0.05,0.03,0.01$). It is obvious that the estimated excess noise is decreasing when the value of $M$ is increasing, which indicates that, in a practical CVQKD system, Alice and Bob will underestimate the excess noise under the effect of the induced photorefraction, and the induced photorefraction may be utilized by the eavesdropper Eve to conceal her eavesdropping behavior without being detected. Subsequently, we will use the classical full intercept-resend attack as an example to further analyze the security of a practical CVQKD system under the effect of the induced photorefraction.

The intercept-resend attack is a well-established attack scheme and its security analysis has been studied in previous work \cite{r34}. In this scheme, Eve intercepts all the pulses transmitted on the quantum channel and uses a heterodyne detector to simultaneously measure the quadratures $X$ and $P$ of the transmitted Gaussian-modulated coherent states. Eve then prepares new coherent states based on her measurements and sends them to Bob. Due to the measurement disturbance and coherent-state shot noise, the full intercept-resend attack will introduce two shot-noise units of excess noise \cite{r17}. If Alice and Bob detect the excess noise introduced by the full intercept-resend attack through parameter estimation, they will stop the CVQKD process immediately to secure the system.

According to Eq. (\ref{e19}) and above analysis, we can get
\begin{equation}\label{e20}
    \begin{split}
        {\varepsilon}_{pra}^{F\!I\!R\!A}&=\varepsilon_{pra}+2,\\
        {\varepsilon}_{est}^{F\!I\!R\!A}&=\frac{{\varepsilon}_{pra}+2}{M},
    \end{split}
\end{equation}
where ${\varepsilon}_{pra}^{F\!I\!R\!A}$ and ${\varepsilon}_{est}^{F\!I\!R\!A}$ are the practical excess noise and estimated excess noise when Eve performs the full intercept-resend attack with the help of the induced photorefraction. When performing the full intercept-resend attack in a practical CVQKD system, we assume the practical excess noise is a typical value, i.e., $\varepsilon_{pra}=0.1$. Therefore, the Eq. (\ref{e20}) becomes
\begin{equation}\label{e21}
    \begin{split}
        {\varepsilon}_{pra}^{F\!I\!R\!A}&=0.1+2=2.1,\\
        {\varepsilon}_{est}^{F\!I\!R\!A}&=\frac{2.1}{M}.
    \end{split}
\end{equation}
If the value of $M$ is equal to $21$, we can obtain
\begin{equation}\label{e22}
    {\varepsilon}_{est}^{F\!I\!R\!A}=\frac{2.1}{21}=0.1={\varepsilon}_{pra}.
\end{equation}

Based on Eq. (\ref{e21}), we can conclude that when the value of $M$ increases, the value of estimated excess noise ${\varepsilon}_{est}^{F\!I\!R\!A}$ will decrease. Therefore, Alice and Bob will underestimate the value of the excess noise under the full intercept-resend attack with the help of the induced photorefraction. According to Eq. (\ref{e22}), it is worth noting that the estimated excess noise ${\varepsilon}_{est}^{F\!I\!R\!A}$ under the full intercept-resend attack can be equal to the practical excess noise $\varepsilon_{pra}$ under no attack, which further indicates that the underestimated excess noise can create a condition for Eve to successfully conceal her eavesdropping behavior in a practical CVQKD system.
\subsection{The scheme of induced-photorefraction attack}\label{subsec3-2}
\begin{figure}[!h]\center
\centering
\resizebox{8cm}{!}{
    \includegraphics{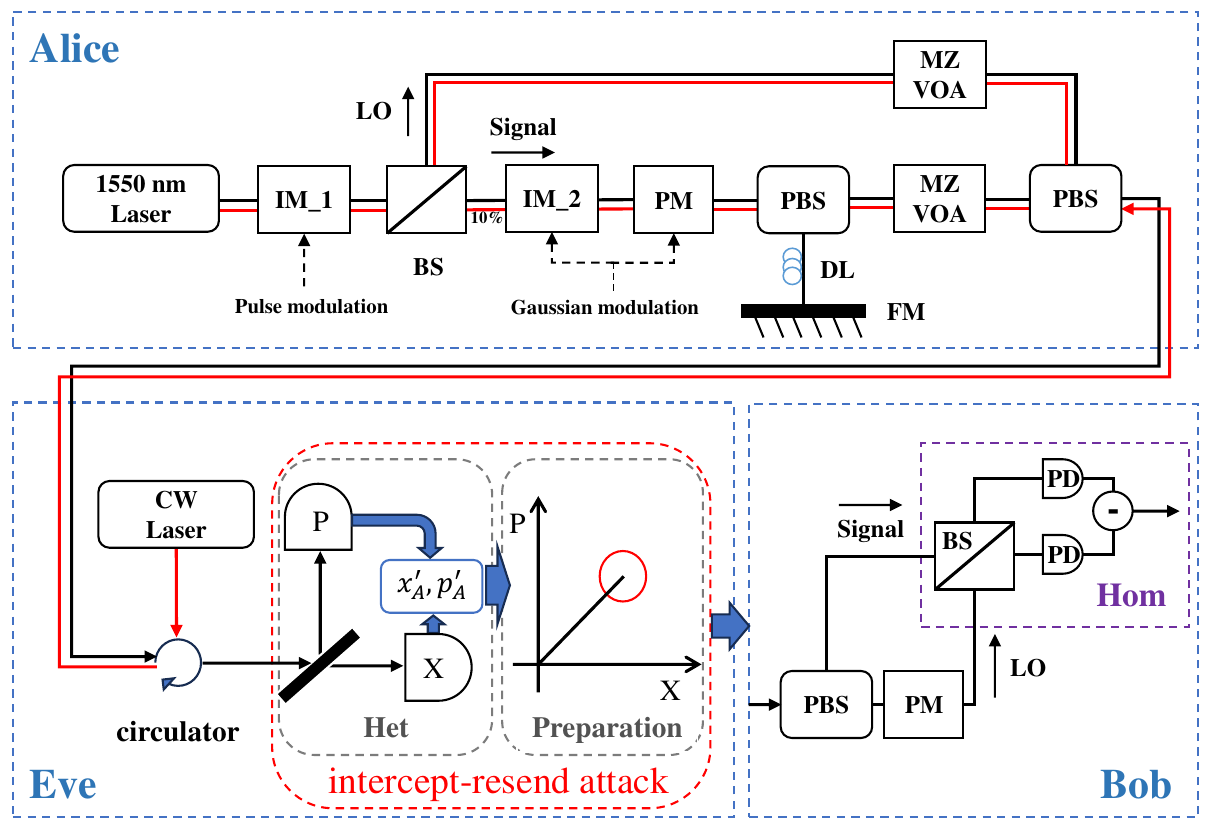}}
\caption{Diagram illustrating the induced-photorefraction attack in the GMCS CVQKD system.}
\label{FIG8}
\end{figure}

Based on the above analysis, Eve can actively open a security loophole to successfully obtain the secret key information by utilizing the full intercept-resend attack with the help of the induced photorefraction, which is an effective quantum hacking strategy, i.e., induced-photorefraction attack (IPA).  Figure \ref{FIG8} shows that the process of the IPA in a practical GMCS CVQKD system, and illustrates that IPA consists of two parts: the induced photorefraction and the full intercept-resend attack. Since the scheme of the full intercept-resend attack has been introduced in Section \ref{subsec3-1}, we will mainly focus on the schemes of the induced photorefraction in the following analysis.

The first scheme of the induced photorefraction utilized by IPA is called pretreatment technique. Specifically, by injecting the irradiation beam at the continuous-wave mode and loading different modulation voltages, we can shift the response curve of a LN-based MZ modulator to any position of the entire $2\pi$ range \cite{r35}. Therefore, Eve can preset the initial state of the modulator to the most beneficial state for IPA. Furthermore, because the PE in LN can always last for a long time \cite{r32,r36}, the pretreatment results can remain in the LN devices for a sufficient time, which will affect their operating status in the CVQKD system.

Because PE is related to the duty cycle, period, and frequency of the irradiation pulse, Eve can use the pulse-injection technique to achieve more accurate and arbitrarily controllable IPA. This technique is divided into two phases, the injection phase and the stabilization phase. In the injection phase, the irradiation beam at the pulse mode is first injected into the LN-based MZ modulator to change the refractive index. Eve can then manipulate the effect of the attack with arbitrary precision by controlling the duty cycle, period, and frequency of the irradiation pulse. In the stabilization phase, when the desired level of attack is reached, Eve can stabilize the effect of the attack by decreasing the duty cycle of the pulse, which essentially counteracts the decay behavior of the PE \cite{r37}.
\subsection{Secret key rate under the induced-photorefraction attack}\label{subsec3-3}
In the above analysis, we mainly focus on the fact that Eve can utilize the IPA to arbitrarily manipulate the parameter estimation of the transmissivity $T$ and excess noise $\varepsilon$ in order to successfully conceal her intercept-resend behavior with the help of the induced photorefraction, which seriously threatens the practical security of the CVQKD system. To show the above effects more clearly, we will further discuss the practical security of the system by analyzing the secret key rate under IPA in the following. The calculation of the secret key rate for the CVQKD system is shown in Appendix \ref{appendixB}. Based on the above calculations and analysis, it can be concluded that the secret key rate with finite-size effect against collective attacks can be regarded as a function of the involved parameters, i.e., $K=K(V_A,T,\varepsilon,\eta,v_{el})$. According to Eqs. (\ref{e18}) and (\ref{e19}), the induced-photorefraction attack will have an effect on the estimated values of the channel parameters $T$ and $\varepsilon$. In addition, based on the Eq. (\ref{e12}), when Alice and Bob do not monitor the modulation variance, they cannot perceive the change of $V_A$. Therefore, when Eve performs the induced-photorefraction attack in a practical CVQKD system, Alice and Bob will use the estimated channel parameters $T_{est}$ and $\varepsilon_{est}$, and the initial modulation variance $V_A$ to calculate the estimated secret key rate, which can be expressed as $K_{est}=K(V_A,T_{est},\varepsilon_{est},\eta,v_{el})$. However, the practical secret key rate should be written as $K_{pra}=K(V_{A}^{\prime},T_{pra},\varepsilon_{pra},\eta,v_{el})$. In order to clearly demonstrate the difference between the estimated secret key rate and the practical secret key rate, we simulate the relationship between the secret key rate and the transmission distance under different intensities of the induced-photorefraction attack and different excess noises. It is noteworthy that the value of $M$ mentioned in Eq. (\ref{e10}) reflects the intensity of the induced-photorefraction attack. And the fixed parameters for the simulation are set as $V_{\!A}=4, \eta=0.5, v_{el}=0.01, T=10^{{-\alpha L}/10}, \alpha=0.2{dB}/{km}, \beta=0.95, N=10^9, n=0.5\times N$, and $\bar{\epsilon}=\epsilon_{P\!A}=10^{-10}.$
\begin{figure}[!h]\center
\centering
\resizebox{8cm}{!}{
    \includegraphics{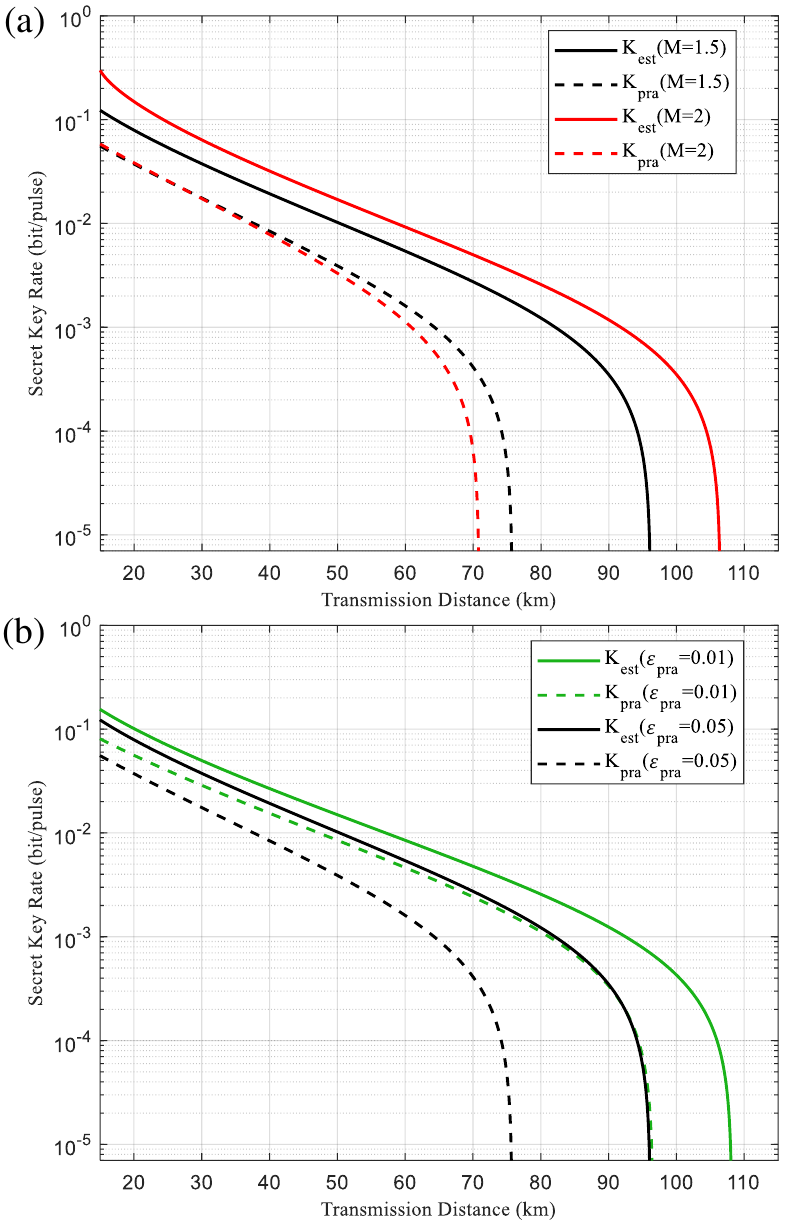}}
\caption{Secret key rate versus transmission distance under (a) different intensities of the induced-photorefraction attack when $\varepsilon_{pra}=0.05$ and (b) different excess noises when $M=1.5$.}
\label{FIG9}
\end{figure}

Figure {\ref{FIG9}}(a) depicts the relationship between the secret key rate and the transmission distance under different intensities of the induced-photorefraction attack (i.e., $M=1.5,2$) when $\varepsilon_{pra}\!=\!0.05$. Figure {\ref{FIG9}}(b) demonstrates the secret key rate versus the transmission distance under different excess noises (i.e., $\varepsilon_{pra}\!=\!0.01,0.05$) when $M=1.5$. These simulation results illustrate that the estimated secret key rate $K_{est}$ is overestimated compared with the practical secret key rate $K_{pra}$ when Eve performs the induced-photorefraction attack. And the overestimation of the secret key rate indicates that the induced-photorefraction attack can open a security loophole for Eve to acquire the secret key information without being detected in a practical CVQKD system.

\begin{figure}[!h]\center
\centering
\resizebox{8cm}{!}{
    \includegraphics{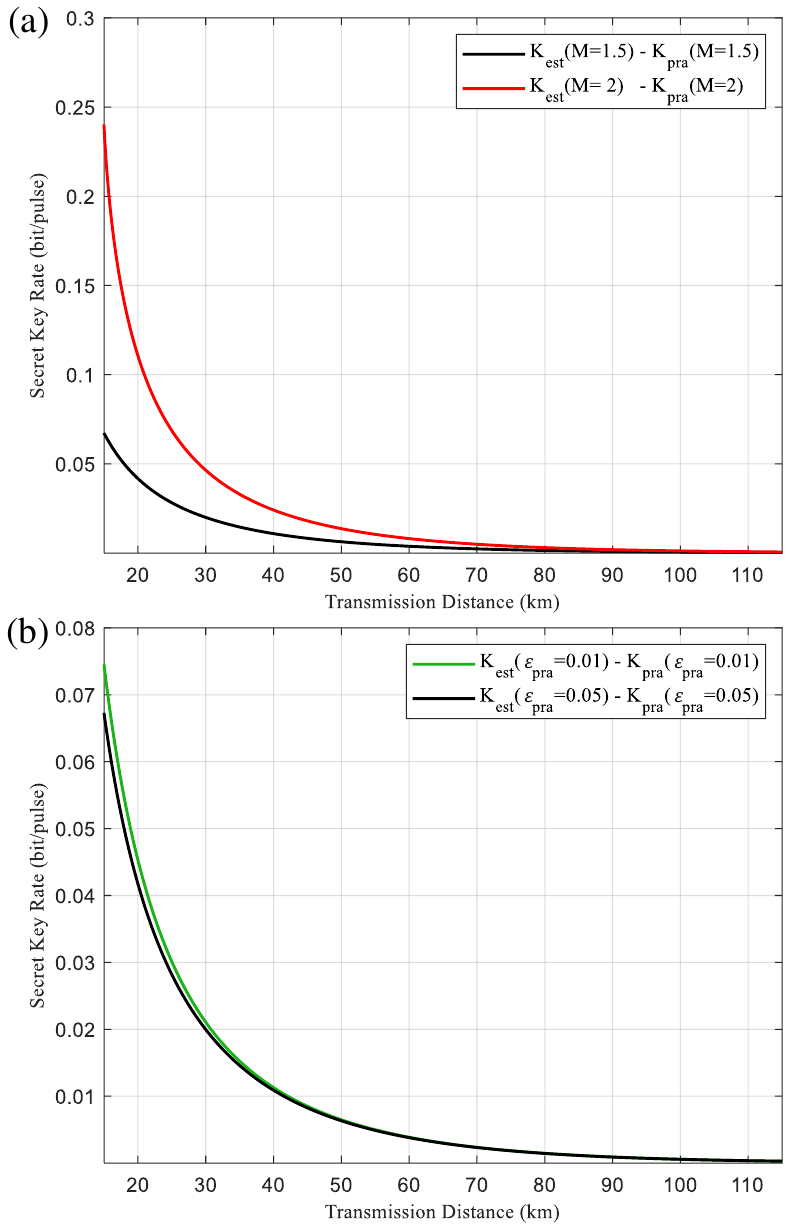}}
\caption{Difference between the estimated secret key rate $K_{est}$ and the practical secret key rate $K_{pra}$.}
\label{FIG12}
\end{figure}

It should be noted that the difference between the estimated secret key rate $K_{est}$ and the practical secret key rate $K_{pra}$ reflects the secret key information that can be obtained by Eve through the intercept-resend attack. In order to clearly show the difference between the estimated secret key rate $K_{est}$ and the practical secret key rate $K_{pra}$, we simulate the values of $K_{est}-K_{pra}$ under different intensities of the induced-photorefraction attack and different excess noises in Fig. \ref{FIG12}. Figure \ref{FIG12}(a) shows that, in a practical CVQKD system, the more serious the induced-photorefraction attack (i.e., the larger the value of $M$), the more secret key information Eve can acquire. In addition, it is obvious that Eve can obtain more secret key information when the excess noise is smaller and the induced-photorefraction attack is of same intensity, which is shown in Fig. \ref{FIG12}(b).
\section{Countermeasures}\label{sec4}
The above investigations indicate that Eve can utilize the induced-photorefraction attack to affect the channel parameter estimation and the calculation of the secret key rate. And these effects open the corresponding security loophole for Eve to successfully perform the full intercept-resend attack and obtain the secret key information in a practical CVQKD system. Therefore, it is necessary to close this security loophole to ensure the security of the system.

\begin{figure}[!h]\center
    \centering
    \resizebox{8cm}{!}{
        \includegraphics{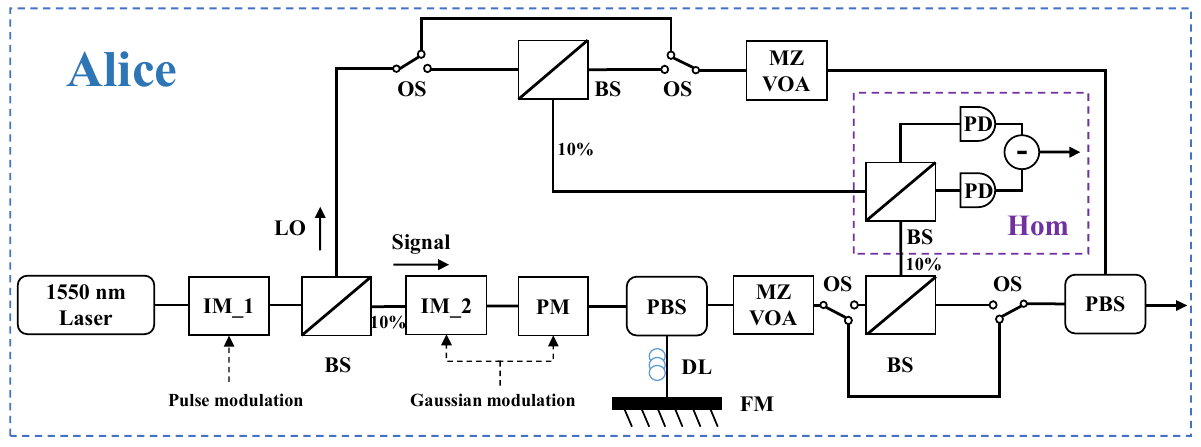}}
    \caption{A random monitoring scheme of the induced-photorefraction attack for a CVQKD system. OS, optical switch.}
    \label{FIG10}
\end{figure}

In a practical CVQKD system, a monitoring scheme is one of the most common approaches to close the security loophole induced by imperfections in practical devices. Based on Eq. (\ref{e12}), we can obtain $M={V_A^\prime}/{V_A}$. This indicates that the change of the modulation variance $V_A$ can reflect the intensity of the induced-photorefraction attack. Therefore, we design a random monitoring scheme of modulation variance for the CVQKD system (see Fig. \ref{FIG10}). Specifically, Alice first randomly utilizes the optical switch and beam splitter to separate a fraction of the undiminished LO light and the attenuated signal light. Here, we minimize the effect of the additional beam splitter on the signal light by using optical switches and beam splitters to achieve random monitoring. Then, the separated signal and LO light interfere in a homodyne detector. Lastly, based on the interference results, Alice can calculate the value of $M$ to evaluate the intensity of the induced-photorefraction attack. Accordingly, we propose using a random monitoring scheme of modulation variance to close the security loophole caused by the induced-photorefraction attack in a practical CVQKD system.

Furthermore, because the first step of the induced-photorefraction attack is to inversely inject the irradiation beam into the LN-based MZ modulators, we can employ isolators or circulators to effectively isolate the irradiation beam from the channel. However, researchers recently show that the security of these devices can also be damaged by Eve in a practical QKD system \cite{r38,r39}. Fortunately, we note that the Ref. \cite{r40} recently proposed an improved optical power limiter and named it the OPL-T. The proposed OPL-T based on the thermal defocusing effect of acrylic prisms can be utilized to limit the irradiation beam and detect Eve’s attacks by monitoring the temperature. Moreover, the limitation performance of the OPL-T has been validated through the experimental demonstration. Therefore, the OPL-T can be used to effectively limit the irradiation beam and reduce the requirement for the isolators and circulators, which will improve the practical security of the CVQKD system.

Besides, in a practical CVQKD system, LN-based MZ IM is the mainstream, although we can use other types of VOA instead of LN-based MZ VOA. Here, we propose to enhance the stability of the IM by replacing the LN-based MZ IM with a Sagnac-based IM.
\begin{figure}[!h]\center
    \centering
    \resizebox{8cm}{!}{
        \includegraphics{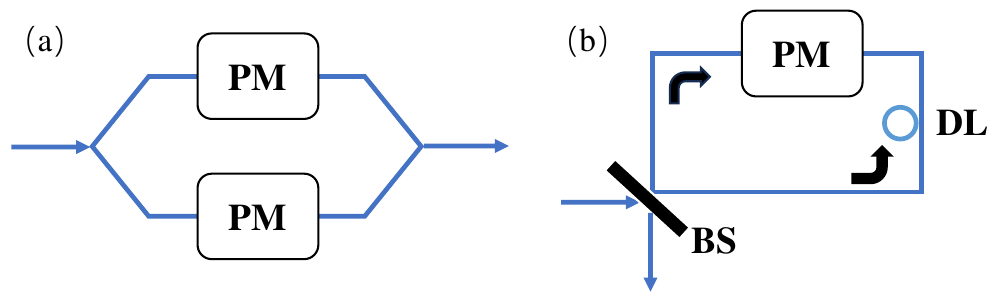}}
    \caption{Intensity modulation. (a) Schematic of the LN-based MZ IM. (b) Schematic of the Sagnac-based IM.}
    \label{FIG11}
\end{figure}
Figure {\ref{FIG11}}(a) shows the simplified structure schematic of the LN-based MZ IM (see Fig. \ref{FIG3} for details). Figure {\ref{FIG11}}(b) illustrates that the Sagnac-based IM consists mainly of a beam splitter, a phase modulator, and a delay line. The work principle of the Sagnac-based IM can be described as follows: when the signal light enters the Sagnac-based IM, it is first divided into two beams by the beam splitter. Subsequently, due to the existence of the delay line, these two beams will pass through the phase modulator at different times. Finally, these two beams will reach the beam splitter at the same time and then interfere to achieve the intensity modulation. Figure \ref{FIG11} depicts the comparison diagram between the LN-based MZ IM and the Sagnac-based IM. The major difference from the LN-based MZ IM is that the two beams in the Sagnac-based IM have the same optical path. This means that any perturbations to the intensity modulation affect both beams equally, such as the fabrication error and the applied electric fields mentioned in Section \ref{subsec2-2}. Therefore, the Sagnac-based IM has natural stability, which can minimize the effects from the induced-photorefraction attack. More importantly, through combining the above countermeasures, legitimate communication parties can accurately calculate the secret key rate of the practical CVQKD system, which can effectively defend against the induced-photorefraction attack.
\section{Conclusion}\label{sec5}
We propose an induced-photorefraction attack and investigate the effects of this attack on the GMCS CVQKD system. We reveal that the response curve of a LN-based MZ modulator will drift when Eve performs this attack, which may amplify the intensity of the optical signal sent by Alice. Furthermore, we analyze the parameter estimation and the secret key rate of the system under the induced-photorefraction attack. The results show that with the help of this attack, the estimated excess noise will be underestimated and the secret key rate of the system will be overestimated, which will open a security loophole for Eve to successfully perform the full intercept-resend attack in a practical CVQKD system. We find that Eve can obtain more secret key information when the induced-photorefraction attack is more serious. In addition, when the intensity of the induced-photorefraction attack is the same, the smaller the channel excess noise is, the more secret key information Eve can acquire. In order to close the security loophole caused by the induced-photorefraction attack, we propose a random monitoring scheme for the intensity of this attack by calculating the change of the modulation variance, and an improving optical power limiter named OPL-T to effectively limit the irradiation beam. Moreover, we also propose using the Sagnac-based IM to replace the LN-based MZ IM for enhanced stability, which can minimize the effects from the induced-photorefraction attack.

\section*{Acknowledgments}
This work was supported by the Joint Funds of the National Natural Science Foundation of China under grant No. U22B2025, Key Research and Development Program of Shaanxi under grant No. 2024GX-YBXM-077, and the Fundamental Research Funds for the Central Universities under grant No. D5000210764.

\appendix
\section{GMCS CVQKD SYSTEM AND ENTANGLEMENT-BASED SCHEME}\label{appendixA}
\begin{figure}[!h]\center
    \centering
    \resizebox{8cm}{!}{
        \includegraphics{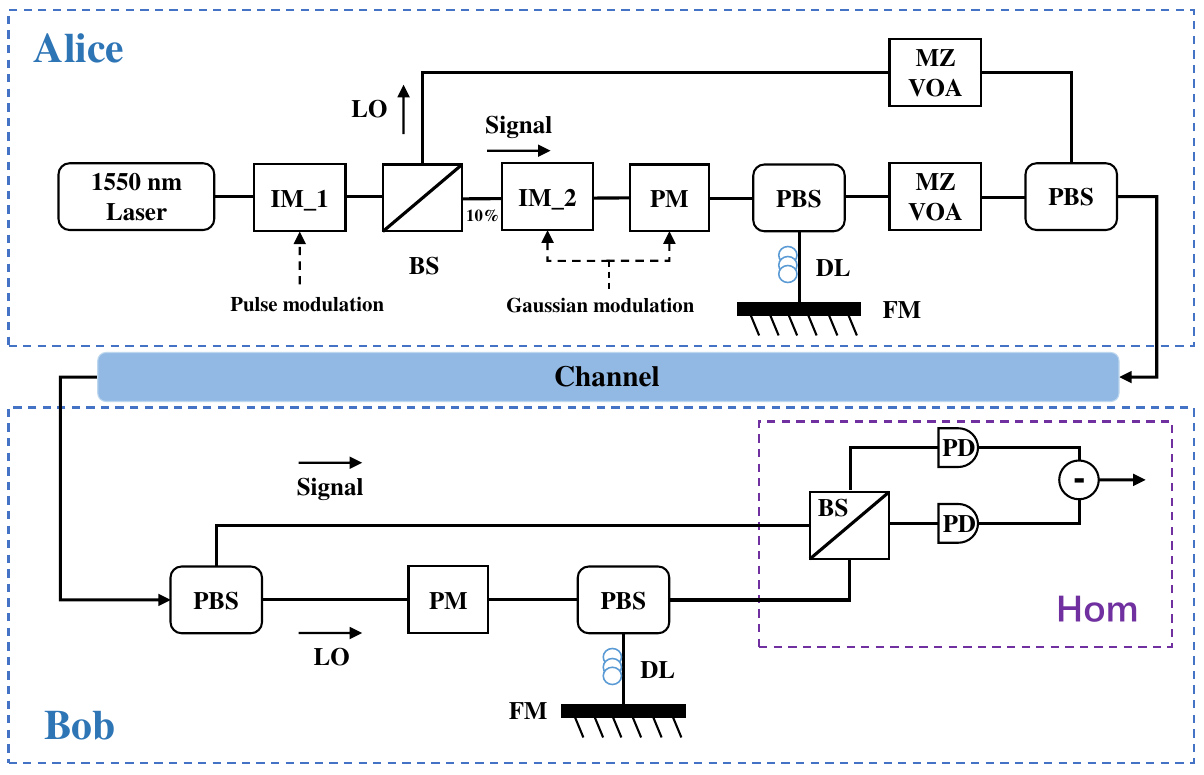}}
    \caption{Practical optical path of the GMCS CVQKD system. IM, intensity modulator; BS, beam splitter; LO, local oscillator; PM, phase modulator; PBS, polarization beam splitter; DL, delay line; FM, Faraday mirror; MZ VOA, Mach-Zehnder interferometer-based variable optical attenuator; Hom, homodyne detector; PD, photodetector.}
    \label{FIG1}
\end{figure}

Figure \ref{FIG1} shows the practical optical path of the GMCS CVQKD system. Specifically, Alice first generates initial coherent states by using a 1550 nm laser and pulse modulation. Subsequently, the initial coherent states are separated into a signal optical path and a LO optical path by a beam splitter. In the signal optical path, Alice encodes secret key information on the quadratures $X$ and $P$ of coherent states with a centered bivariate Gaussian modulation. After attenuation by a variable optical attenuator, the signal light and LO light are transmitted to Bob in the same quantum channel by time-division multiplexing and polarization multiplexing. On the receiver side, Bob selects the measurement basis by varying the phase difference between the signal light and LO light, and randomly measures the quadrature $X$ or $P$ by performing a balanced homodyne detection. In addition to the above steps, Alice and Bob need to perform the classical data post-processing operations to share the final secret key, including parameter estimation, reverse reconciliation, and privacy amplification.
\begin{figure}[!h]\center
    \centering
    \resizebox{8cm}{!}{
        \includegraphics{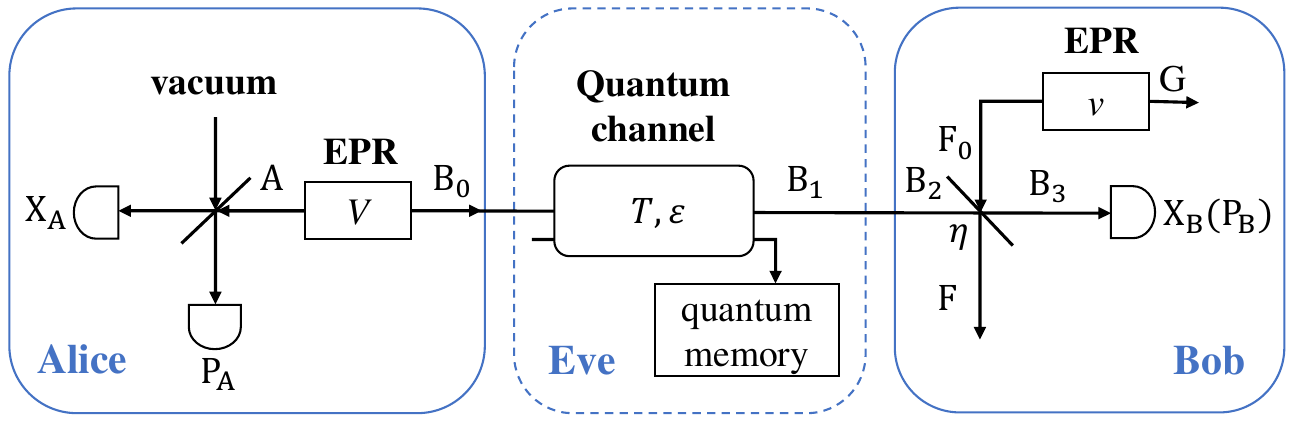}}
    \caption{Entanglement-based scheme of a GMCS CVQKD system with homodyne detection. The transmissivity $T$ and excess noise $\varepsilon$ of the quantum channel can be controlled by Eve.}
    \label{FIG2}
\end{figure}

The entanglement-based (EB) scheme is commonly used to evaluate the security of the practical GMCS CVQKD systems (see Fig. \ref{FIG2}). In this scheme, Alice models the coherent state preparation using a two-mode squeezed vacuum (EPR) state with variance $V= V_A+1$, where $V_A$ is Alice’s modulation variance. Half of the EPR state is measured by Alice using a heterodyne detector and the other half is sent to Bob through a quantum channel with transmissivity $T$ and excess noise $\varepsilon$. When Bob receives the mode $B_2$, he measures the quadrature $X$ or $P$ by using a homodyne detector with efficiency $\eta$ and electronic noise $v_{el}$. Here, the inefficiency of the detector is modeled by a beam splitter with transmission $\eta$, and the electronic noise is modeled by an EPR state with variance $v$, where $v=1+v_{el}/(1-\eta)$. 

\section{SECRET KEY RATE OF THE GMCS CVQKD SYSTEM}\label{appendixB}
After parameter estimation, Alice and Bob will utilize $n$ received pulses to establish the secret key of the GMCS CVQKD system. In the case of collective attacks, the theoretical secret key rate of the system considering the reverse reconciliation and finite-size effect can be expressed as \cite{r8,r41}
\begin{equation}\label{e23}
    K=\frac{n}{N}[\beta I_{A\!B}-S_{B\!E}^{\epsilon_{P\!E}}-\Delta(n)],\\
\end{equation}
where $n=N-m$ and $\beta$ is the reverse reconciliation efficiency. The mutual information $I_{AB}$ between Alice and Bob can be represented as
\begin{equation}\label{e24}
    I_{A\!B}=\log_2{\frac{V_B}{V_{B|A}}}=\log_2{\frac{V+\chi_{tot}}{1+\chi_{tot}}},\\
\end{equation}
where $V=V_A+1$, $\chi_{tot}=\chi_{line}+\chi_{hom}/T$ is the total noise referred to the channel input, $\chi_{line}=1/T-1+\varepsilon$ is the total channel-added noise referred to the channel input, and $\chi_{hom}=[(1-\eta)+v_{el}]/\eta$ is the detection-added noise referred to Bob’s input. $S_{B\!E}^{\epsilon_{P\!E}}$ is the maximum value of the Holevo information compatible with the statistics except with probability $\epsilon_{P\!E}$. In particular, the covariance matrix between Alice and Bob is related to $S_{B\!E}^{\epsilon_{P\!E}}$, which can be calculated as
\begin{equation}\label{e25}
    \begin{split}
        \Upsilon_{A\!B}&=
        \left[\begin{array}{ccc}
            \Upsilon_A & \sigma_{A\!B}^T\\
            \sigma_{A\!B} &\Upsilon_{B}
        \end{array}
        \right]\\
        &=
        \left[\begin{array}{ccc}
            VI_2 & \sqrt{T_{min}(V^2-1)}\sigma_z\\
            \sqrt{T_{min}(V^2-1)}\sigma_z & [T_{min}(V+\chi_{line,max})]I_2
        \end{array}
        \right],
    \end{split}
\end{equation}
where $I_2=diag[1,1]$, $\sigma_z=diag[1,-1]$, $\chi_{line,max}=1/{T_{min}}-1+\varepsilon_{max}$, and $T_{min}$ and $\varepsilon_{max}$ correspond to the lower bound of $T$ and the upper bound of $\varepsilon$, respectively.

Based on the analysis in Section \ref{subsec3-1}, the quantum channel involved in a CVQKD system is assumed to be a linear model and the parameters $T$ and $\xi$ are estimated by using $m$ pairs data from the $X$ or $P$. When $m$ is large enough (e.g., $m>10^6$), $T_{min}$ and $\varepsilon_{max}$ can be calculated as \cite{r41}
\begin{equation}\label{e26}
    \begin{split}
        T_{min}&=\frac{(\hat{t}-\Delta t)^2}{\eta},\\
        \varepsilon_{max}&=\frac{{\hat{\sigma}}^2+\Delta {\sigma}^2-N_0-v_{el}N_0}{{\hat{t}}^2N_0}.
    \end{split}
\end{equation}
For a linear model, the maximum-likelihood estimators $\hat{t}$ and ${\hat{\sigma}}^2$ can be expressed as
\begin{equation}\label{e27}
    \hat{t}=\frac{\sum_{i=1}^{m}{x_A}_i{x_B}_i}{\sum_{i=1}^{m}{{x^2_A}_i}},\qquad{\hat{\sigma}}^2=\frac{1}{m}\sum_{i=1}^{m}({x_B}_i-\hat{t}{x_A}_i)^2.\\
\end{equation}
In addition, $\Delta t$ and $\Delta{\sigma}^2$ can be calculated as
\begin{equation}\label{e28}
    \Delta t=z_{\epsilon_{P\!E}/2}\sqrt{\frac{{\hat{\sigma}}^2}{mV_{x_A}}},\qquad\Delta{\sigma}^2=z_{\epsilon_{P\!E}/2}\frac{{\hat{\sigma}}^2\sqrt{2}}{\sqrt{m}}.\\
\end{equation}
where the coefficient $z_{\epsilon_{P\!E}/2}$ satisfies the following relation:
    $1-\frac{1}{2}er\!f(z_{\epsilon_{P\!E}/2}/\sqrt{2})={\epsilon_{P\!E}}/2$, and $er\!f(\cdot)$ is the error function which can be expressed as $er\!f (x)=2\pi^{-\frac{1}{2}}\int_{0}^{x}e^{-t^2}dt$.

Then $S_{B\!E}^{\epsilon_{P\!E}}$ can be calculated as
\begin{equation}\label{e29}
    S_{B\!E}^{\epsilon_{P\!E}}=\sum_{i=1}^{2}G\Big(\frac{\lambda_i-1}{2}\Big)-\sum_{i=3}^{5}G\Big(\frac{\lambda_i-1}{2}\Big),\\
\end{equation}
where $G(x)=(x+1)\log_2(x+1)-x\log_2x$ and $\lambda_i$ are the symplectic eigenvalues of the covariance matrix between Alice and Bob, which can be expressed as
\begin{equation}\label{e30}
    \begin{split}
        \lambda_{1,2}^{2}&=\frac{1}{2}(A\pm\sqrt{A^2-4B}),\\
        \lambda_{3,4}^{2}&=\frac{1}{2}(C\pm\sqrt{C^2-4D}),\\
        \lambda_5&=1.
    \end{split}
\end{equation}
Here
\begin{equation}
    \label{e31}
    \begin{split}
        A=&det\Upsilon_{\!A}\!+\!det\Upsilon_{\!B}\!+\!2det\sigma_{\!A\!B}\\
        \quad=&V^2(1-2T_{min})+2T_{min}+{T_{min}}^2(V+\chi_{line,max})^2,\\
        B=&det\Upsilon_{\!A\!B}={T_{min}}^2(V\chi_{line,max}+1)^2,\\
        C=&\frac{A\chi_{hom}+V\sqrt{B}+T_{min}(V+\chi_{line,max})}{T_{min}(V+\chi_{line,max}+\chi_{hom}/T_{min})},\\
        D=&\sqrt{B}\frac{V+\sqrt{B}\chi_{hom}}{T_{min}(V+\chi_{line,max}+\chi_{hom}/T_{min})}.
    \end{split}
\end{equation}

Furthermore, in a practical CVQKD system, $\Delta(n)$ is related to the security of the privacy amplification, which can be written as
\begin{equation}
    \label{e32}
    \Delta(n)=7\sqrt{\frac{\log_2(1/\bar{\epsilon})}{n}}+\frac{2}{n}\log_2{\frac{1}{\epsilon_{P\!A}}},\\
\end{equation}
where $\bar{\epsilon}$ and $\epsilon_{P\!A}$ represent the smoothing parameter and the failure probability of privacy amplification, respectively. And because the value of $\Delta(n)$ mainly depends on $n$, the values of $\bar{\epsilon}$ and $\epsilon_{P\!A}$ are usually set to be equal to the value of $\epsilon_{P\!E}$.

\bibliographystyle{unsrt}
\bibliography{Quantum_hacking_Induced-photorefraction_attack_on_a_practical_continuous-variable_quantum_key_distribution_system.bib}
	
\end{document}